\documentclass[manuscript]{aastex}

\usepackage{txfonts}

\usepackage{graphicx}
\usepackage{epstopdf}

\usepackage{longtable}

\usepackage{natbib}

\slugcomment{For submission to the Astrophysical Journal}

\shorttitle{Grid resolution and coronal density}
\shortauthors{Bradshaw and Cargill}

\begin{document}

\title{The influence of numerical resolution on coronal density in hydrodynamic models of impulsive heating}

\author{S. J. Bradshaw}
\affil{Department of Physics and Astronomy, Rice University, Houston, TX 77005, USA}
\email{stephen.bradshaw@rice.edu}\and
\author{P. J. Cargill}
\affil{Space and Atmospheric Physics, The Blackett Laboratory, Imperial College, London, SW7 2BW, UK \and School of Mathematics and Statistics, University of St Andrews, St Andrews, Scotland, KY16 9SS, UK}
\email{p.cargill@imperial.ac.uk}

\begin{abstract}
The effect of the numerical spatial resolution in models of the solar corona and corona~/~chromosphere interface is examined for impulsive heating over a range of magnitudes using one dimensional hydrodynamic simulations. It is demonstrated that the principle effect of inadequate resolution is on the coronal density. An under-resolved loop typically has a peak density of {\it at least} a factor of two lower than a resolved loop subject to the same heating, with larger discrepencies in the decay phase. The temperature for under-resolved loops is also lower indicating that lack of resolution does not ``bottle up'' the heat flux in the corona. Energy is conserved in the models to under 1\% in all cases, indicating that this is not responsible for the low density. Instead, we argue that in under-resolved loops the heat flux ``jumps across'' the transition region to the dense chromosphere from which it is radiated rather than heating and ablating transition region plasma. This emphasises the point that the interaction between corona and chromosphere occurs only through the medium of the transition region. Implications for three dimensional magnetohydrodynamic coronal models are discussed.
\end{abstract}


\section{Introduction}
\label{introduction}

The development of multi-dimensional computational models of the solar corona is rendered difficult because of the variety of physical processes operating over a wide range of spatial and temporal scales. Of particular importance are the very different requirements for studying the evolution and dissipation of magnetic fields, and the plasma response to any particle heating and acceleration. The former requires multi-dimensional (e.g. 3D) magnetohydrodynamic (MHD) models, often with a simple energy equation, while the latter usually involves the solution of the system of one-dimensional (1D) hydrodynamic (HD) equations for a prescribed magnetic field geometry, with an energy equation that aims to capture all of the important plasma processes, and also includes detailed spectral modeling for direct comparison with observations. Combining these two aspects into a unified approach is an ongoing challenge \citep[e.g.][]{Mikic_1999,Peter_2004,Gudiksen_2005a,Gudiksen_2005b,Peter_2006,Bingert_2011,Gudiksen_2011,Zacharias_2011}.

Despite the apparent simplicity and tractability of the 1D HD approach, it remains a far from trivial computational problem. The treatment of shocks needs care, but is well understood. Despite early recognition of the problems \citep[e.g.][]{Craig_1982}, the numerical treatment of thermal conduction, in particular the resolution of steep temperature gradients, remains a serious concern. This can be seen by considering a static equilibrium loop. The energetics near the loop apex is roughly a balance between the heating function and a downward heat flux. Relatively shallow temperature gradients are sufficient to carry away the excess energy from the corona because the coefficient of thermal conduction scales as $T^{5/2}$. For an active region loop of half-length 25 Mm and apex temperature 4~MK, the coronal temperature length scale, defined as $L_T = T / |dT/ds| = \kappa_0 T^{7/2} / F_c$, is a few thousand km, requiring grid cells of order 1000~km for resolution. Here $F_c$ is the heat flux, $s$ a coordinate along the loop and $\kappa_0$ is the thermal conduction coefficient. Below some point in the atmosphere (defined as the top of the transition region), the energy balance is between the downward heat flux and optically thin radiative losses, with the strong radiation requiring a steep temperature gradient (large heat flux) for balance. For the above example, transition region values of $L_T$ may be of order $\le 1000$~m, which require grid cells of $\le 500$~m for resolution. Hotter loops require smaller grid cells and longer loops can be modelled with wider cells. Such resolutions are straightforward in static loop models, either by brute force or especially using thermally structured grids \citep[e.g. Appendix~A of][]{Cargill_2012}.

In a dynamic, evolving corona, two major problems arise. When using an explicit numerical scheme to solve the energy equation, there is the need to ensure stability. The thermal conduction timescale across a grid cell of width $\Delta s$ is of order $\Delta t_c = 4 \times 10^{-10} n \Delta s^2 /T^{5/2}$ and one requires the timestep ($\Delta t$) to satisfy $\Delta t < \frac{1}{2} \mbox{min} (\Delta t_c)$, where the minimum is taken across the entire grid, a condition which holds irrespective of the state of the loop. [The stability restriction can be removed by use of an implicit code, but the physical timescale implied by, for example, a moving conduction front cannot: see Section~\ref{discussion}.] For a coronal plasma of density $10^9$~cm$^{-3}$ heated to $10^6$~K with a cell width of 200 km, $\Delta t$ is of order 0.1~s \citep{Bradshaw_2010b}. For a transition region plasma of density $10^{10}$~cm$^{-3}$ at $10^5$~K and grid 500 m, $\Delta t$ is of order $2 \times 10^{-3}$~s. Temperatures characteristic of a flare obviously decrease these numbers drastically. So, to follow an event accurately for the timescales of interest (minutes or hours) still requires the resolution of the principle timescales of the physical processes driving the evolution: of order milliseconds or less.

The second problem is that the location of the steepest temperature gradients moves in response to the corona being heated and then cooling. During heating and early cooling phases, a conduction front moves downward in the atmosphere pushing the transition region lower, and during the later cooling the transition region retreats back into the corona. So our 500 m cell mentioned above needs to move around. It is clearly not computatially efficient, or even practical, to use the small-scale grid required by the transition region everywhere in the domain ($\Delta t_c$ in the corona for a 500~m grid cell would be $10^{-6}$~s in the above example), so one needs to ensure  there is high resolution where it is needed at all times, but not over-resolution elsewhere. There are two potential solutions to this problem. The first is to use a fixed but non-uniform grid where the density of grid cells is greatest in regions of the solution that demand the highest resolution. This can be effective if one can be certain that these regions remain more or less stationary and do not propagate to the lower resolution parts of the grid as the equations are advanced in time. In the circumstances under which this cannot be guaranteed then the solution is to use an evolving~/~adaptive grid in which the density of grid cells increases or decreases in different regions of the solution as required. One implementation of this technique keeps the total number of cells constant by stretching~/~shrinking the grid cells, whereas other methods allow a variable number of grid cells \citep[e.g.][]{Betta_1997,MacNeice_2000}.

The question of what constitutes adequate spatial resolution in 1D HD models, and the consequences of not resolving the atmosphere properly, has been of concern for over three decades. One early argument \citep{Craig_1982} was that inadequate resolution would cause the heat flux to become ``bottled up'' in the corona, leading to slower cooling. In this paper we show that this is not the case and explore the issue of real concern, which is how the numerical resolution impacts the interaction between corona, transition region and upper chromosphere, especially the so-called ``chromospheric evaporation'' that provides the corona with density. This has never been quantified and is of enormous interest and importance to coronal modeling, especially in 3D codes where it is harder to achieve adequate resolution. In Section~\ref{numerical} and Appendix~\ref{AppA} the key features of the 1D HD code and the numerical experiments that we carry out are described. In Section~\ref{results} we discuss our findings in relation to the response of the solar atmosphere to heating, mass and energy transport, and the conservation laws. In Section~\ref{discussion} we discuss a number of implications of the results and in Section~\ref{summary} we present our conclusions.

\section{Numerical model and experiments}
\label{numerical}

The effect of numerical resolution is explored using the 1D HYDRAD code \citep{Bradshaw_2011b}, which is discused fully in Appendix~\ref{AppA}, through a series of examples of impulsive coronal heating. The heating covers several orders of magnitude, ranging from that required to maintain quiet Sun conditions to reasonably powerful flares. Short and long loops, and slow and fast heating are considered. The spatial profile of the energy release is uniform along the loop and the temporal profile is triangular, with a linear ramp up to the peak volumetric heating rate followed by a linear decrease to zero over a total period $\tau_H$.

To facilitate this investigation we employ the adaptive regridding capabilities of HYDRAD to study a range of maximum refinement levels. The condition for refinement is such that cell-to-cell changes in the density and electron energy are kept between 5\% and 10\% where possible. The largest grid cell in all of our calculations has a width of $4\times10^7$~cm (400~km) and each refinement splits the cell in two. So, a grid cell that has been refined once has a width of $2\times10^7$~cm, twice yields a width of $10^7$~cm and $RL$~times a width of $4\times10^7 / 2^{RL}$ (where $RL$ is the Refinement Level). We employ a maximum of 12 levels of refinement, corresponding to a grid cell width in the most highly resolved regions of $4\times10^7 / 2^{12}=9.8\times10^3$~cm (98~m). Using the same set of parameter values we repeated each run for $RL=[0,2,4,6,8,10,12]$~refinement levels to create a group of runs. These are summarised in Table~\ref{table1}, where we note that: (1) the total length of the loop ($2L$) includes a 10~Mm chromosphere at each foot-point; and (2) the total energy in the loop at $t=0$ (the initial conditions) is negligible compared with the energy released into the loop during the heating phase.

\begin{longtable}{c c c c c c c c}
\caption{A Summary of the Parameter Space Used to Conduct the Numerical Experiments for $RL=[0,2,4,6,8,10,12]$. The columns show, respectively, the group number, the length of the loop (including a 10 Mm chromosphere at either footpoint), the peak heating rate, the width of the heating pulse, the maximum averaged temperature and density, and the resolution level needed for density to exceed 90 and 75\% of the properly resolved value.}\\
\hline
Group~\# & $2L$ & $E_H$ & $\tau_H$ & $T_{\mbox{max}}$ & $n_{\mbox{max}}$ & 90\% & 75\% \\
 & [Mm] & [erg~cm$^{-3}$~s$^{-1}$] & [s] & [MK] & [10$^9$~cm$^{-3}$] & [RL] & [RL] \\
\hline
\endhead
 1 & 60 & 0.008 & 60 & 3 & 0.6 & 4 & 2 \\
 2 & 60 & 0.08 & 60 & 6.5 & 2 & 8 & 6 \\
 3 & 60 & 0.8 & 60 & 13 & 7 & 10 & 8 \\
 4 & 60 & 0.008 & 600 & 3.5 & 1.6 & 6 & 4 \\
 5 & 60 & 0.08 & 600 & 7 & 8 & 8 & 6 \\
 6 & 60 & 0.8 & 600 & 15 & 35 & 8 & 6 \\
 7 & 180 & 0.0005 & 60 & 2 & 0.3 & 4 & 4 \\
 8 & 180 & 0.005 & 60 & 4 & 0.35 & 6 & 6 \\
 9 & 180 & 0.05 & 60 & 12 & 1 & 8 & 6 \\
10& 180 & 0.0005 & 600 & 3 & 0.3 & 6 & 4 \\
11& 180 & 0.005 & 600 & 6 & 1 & 6 & 4 \\
12& 180 & 0.05 & 600 & 13 & 4 & 8 & 6 \\
\hline
\label{table1}
\end{longtable}

As discussed fully in Appendix~\ref{AppA}, we solve the 1D HD equations appropriate for a single magnetic strand in the field-aligned direction as documented in \cite{Bradshaw_2011b}. In summary, a multi-species approach is adopted by treating electrons and ions as separate fluids, and coupling them via Coulomb collisions. The electrons are heated preferentially ($T_e \neq T_i$), with quasi-neutral ($n_e = n_i = n$) and current free ($v_e = v_i = v$) conditions assumed. The equations solved are formulated to describe the conservation of mass, momentum and energy. This has the advantage of making it particularly easy to determine how well the numerical methods implemented in HYDRAD satisfy the conservation conditions and the role that spatial resolution plays. [We discuss methods that do not solve the conservative form of the energy equation, such as 3D MHD codes, later.] The following physics is included: transport and compression, viscous stress, gravitational acceleration and potential energy, Coulomb collisions, thermal conduction, optically-thin radiation, and external energy input (heating). Collisions between like-species are frequent enough that the electron and ion equations of state are given by $p_e = k_B n T_e$ and $p_i = k_B n T_i$, and only the thermal component of the electron energy and the thermal plus kinetic components of the ion energy are considered. The HYDRAD code, written exclusively in C++, is fast, robust and models an entire loop strand (foot-point to foot-point for any geometry via an analytical equation, or look-up table, for gravity). It is user-friendly and easily configurable via a Java-developed graphical user interface.

\section{Results}
\label{results}

\subsection{Coronal response to heating}
\label{response}

\begin{figure}
\includegraphics[width=0.8\textwidth]{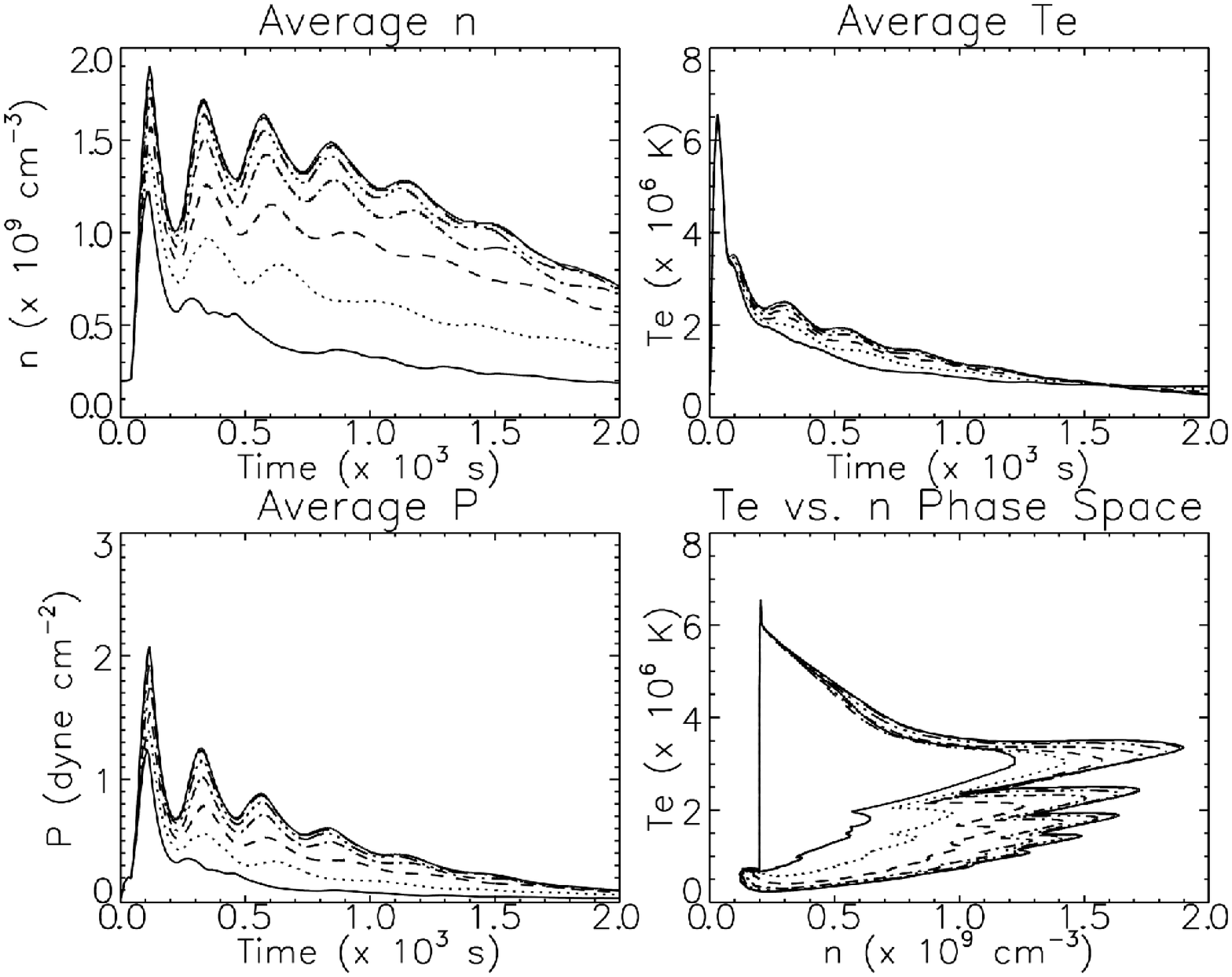}
\includegraphics[width=0.8\textwidth]{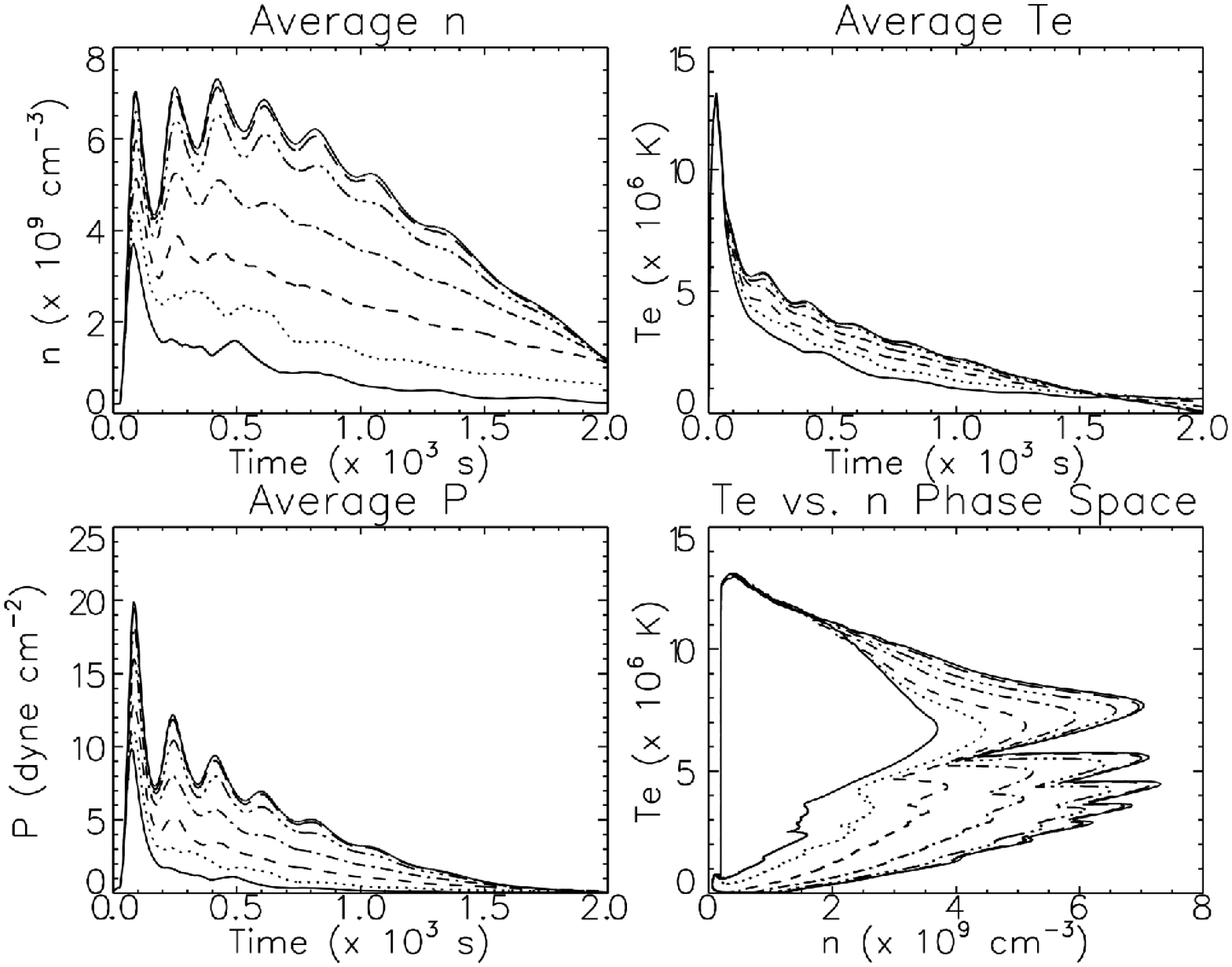}
\caption{Results for groups~2 (upper four panels) and 3 (lower four panels). Starting from the top left, the panels show the average density, electron temperature and pressure as a function of time, and the relation between $T$ and $n$. The various curves represent different values of $RL$, which converge as $RL$ increases. The evolution is clockwise along the curves in the $T_e$ vs. $n$ plots.}
\label{fig1}
\end{figure}

Figure~\ref{fig1} shows the temporal evolution of the spatially averaged density ($n$), electron temperature ($T_e$), total pressure ($P$) and the corresponding $T_e$ vs. $n$ phase-space plot for the numerical runs comprising groups~2 (upper four panels) and 3 (lower four panels), as described in Table~\ref{table1}. These correspond to roughly an active region nanoflare and a small flare respectively. The spatial average was calculated over the upper-most 50\% of the loop, which is comprised solely of coronal plasma. Each curve corresponds to a different maximum refinement level for the adaptive grid. Taking group~2 as an example, the lowest solid line in the upper-left panel for Figure~\ref{fig1} shows the average density for a refinement level of zero (no refinement, a uniform grid of $4\times10^7$~cm grid cells), the dotted line shows the average density for a refinement level of $RL=2$, and so forth. The curves for refinement levels of 10 (long dashes) and 12 (upper solid line) are essentially identical, which indicates that a maximum spatial resolution of order 100~m is required for the density solution to converge for this group of runs. 

It is striking that the average electron temperatures are only weakly dependent on the resolution and converge at $RL=6$: even coarser grids perform adequately. The peak temperature is similar for all resolutions, though under-resolved loops cool somewhat more rapidly than the resolved ones. Since the peak temperature is reached when heat input and coronal energy loss by thermal conduction are in balance \citep{Bradshaw_2006}, one would expect this to require adequate spatial resolution of the temperature gradients. However, we are only considering the average coronal temperatures here and even our coarsest grid is sufficient to resolve coronal scales (of order 1000~km), which is all that is needed for thermal conduction to efficiently remove the excess energy deposited in the corona. This shows that inadequate spatial resolution does not ``bottle up'' the heat flux and thereby slow coronal cooling.

On the other hand the density shows a major difference between low and high resolution. For these two cases with small $\tau_H$ and $L$, the density reaches a peak a few tens of seconds after the temperature and then oscillates as the plasma sloshes to and fro within the loop. For both cases in Figure 1, there is a difference of ~80\% between the initial peaks for $RL = 0$ and $RL = 12$, and a larger difference in the density decay phase, with the poorly resolved cases also showing little in the way of oscillations. Group~2 has fairly moderate heating with a peak electron temperature at just above 6~MK, and yet at 500~s there is a factor of greater than 3 difference in the density between the coarsest and most refined solutions. Even with 25~km grid cells ($RL=4$) the density is about 40\% lower than the converged value. Group~3, for which an order of magnitude more energy was released over the same timescale (60~s) shows greater differences for different values of $RL$ than group~2, though the runs for $RL=[10,12]$ again demonstrate successful convergence. Indeed, this group of runs demonstrates that as one approaches flare territory the demands made of spatial resolution become more extreme and the need to satisfy them to derive accurate solutions becomes more critical. It is evident, for example, that grid cells greater than 1~km in width are not sufficient. The final two columns in Table~\ref{table1} show the resolution required for the density to exceed 90 and 75 \% of its fully resolved value. These numbers should be treated as approximate, and are evaluated around the 2nd and 3rd density peaks. 

The magnitude of the density oscillations are also diminished for poorer resolution, indicative of faster damping. These should damp conductively and we find that while the damping time of the resolved cases is broadly consistent with the calculated conductive damping time, the fast damping of the unresolved cases is not. The conductive timescale scales as $n/T^{5/2}$, but the lower density in the unresolved cases is compensated by a somewhat smaller temperature so that the conductive timescale changes little. So it appears that the lack of resolution also enhances the damping, a result of importance for coronal seismologists who may use 1D HD models to interpret their results \citep[for a further discussion of wave damping in the context of coronal seismology see][]{Morton_2010}.

\begin{figure}
\includegraphics[width=1.0\textwidth]{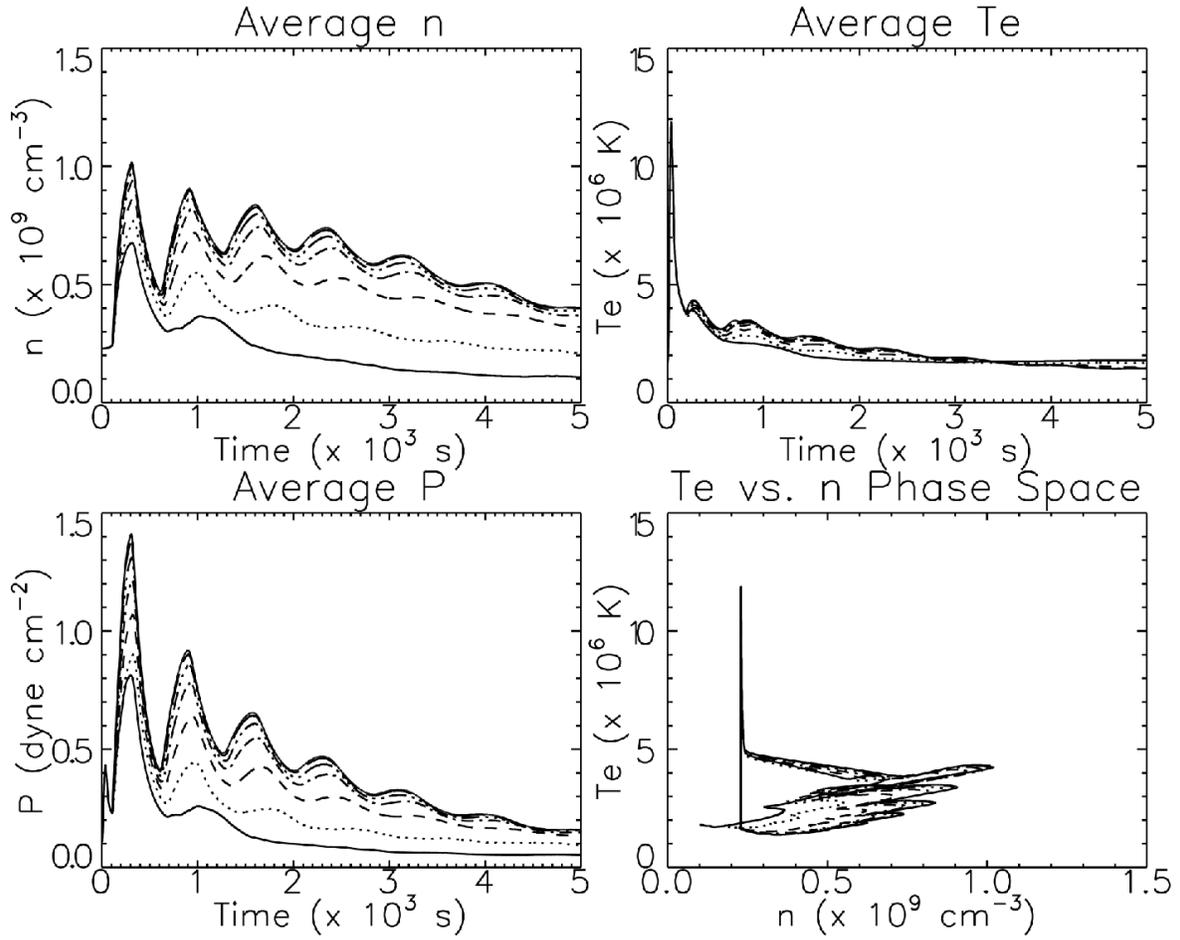}
\caption{Results for group~9. The notation is as in Figure 1}
\label{fig2}
\end{figure}

Figure~\ref{fig2} and Table~\ref{table1} show the set of results for the runs comprising group~9, another very impulsive heating case corresponding to a strong nanoflare in a long (180 Mm) loop. The pattern followed by the density and temperature evolution, and seen in the phase-space plot, is broadly the same as in Figure~\ref{fig1}. However, even though this group is subject to the most extreme volumetric heating rate for the 180~Mm runs, the density difference (e.g. at 1000~s) between the coarsest and most refined grids is only about a factor of 2. The reason for this is that longer loops require shallower temperature gradients and consequently the coarser grids are more equal to the task of providing adequate resolution. For example, $RL=6$ does a reasonable job in this case, though one should bear in mind that this still corresponds to a grid cell width of just over 5~km for a loop of 180~Mm total length, reaching a peak electron temperature in the region of 12~MK. [We note from the phase-space plot that the $RL=[0,2]$ curves do not recover the initial equilibrium, but find another nearby in temperature but of about half the density. The curves for $RL>2$ can all be seen to recover the initial equilibrium. The reason for this is that the initial conditions fed to HYDRAD are calculated on an extremely high resolution grid with cell widths as small as 1~cm; these are then interpolated onto the much coarser grid used for the time-dependent calculation, which is then adaptively refined as the calculation proceeds. The grids of $RL=[0,2]$ are not sufficient to resolve even the initial conditions in the case of group~9 and so these runs ultimately return to a different equilibrium state. The grids of $RL>2$ do not have this problem and converge on the initial equilibrium.]

\begin{figure}
\includegraphics[width=0.8\textwidth]{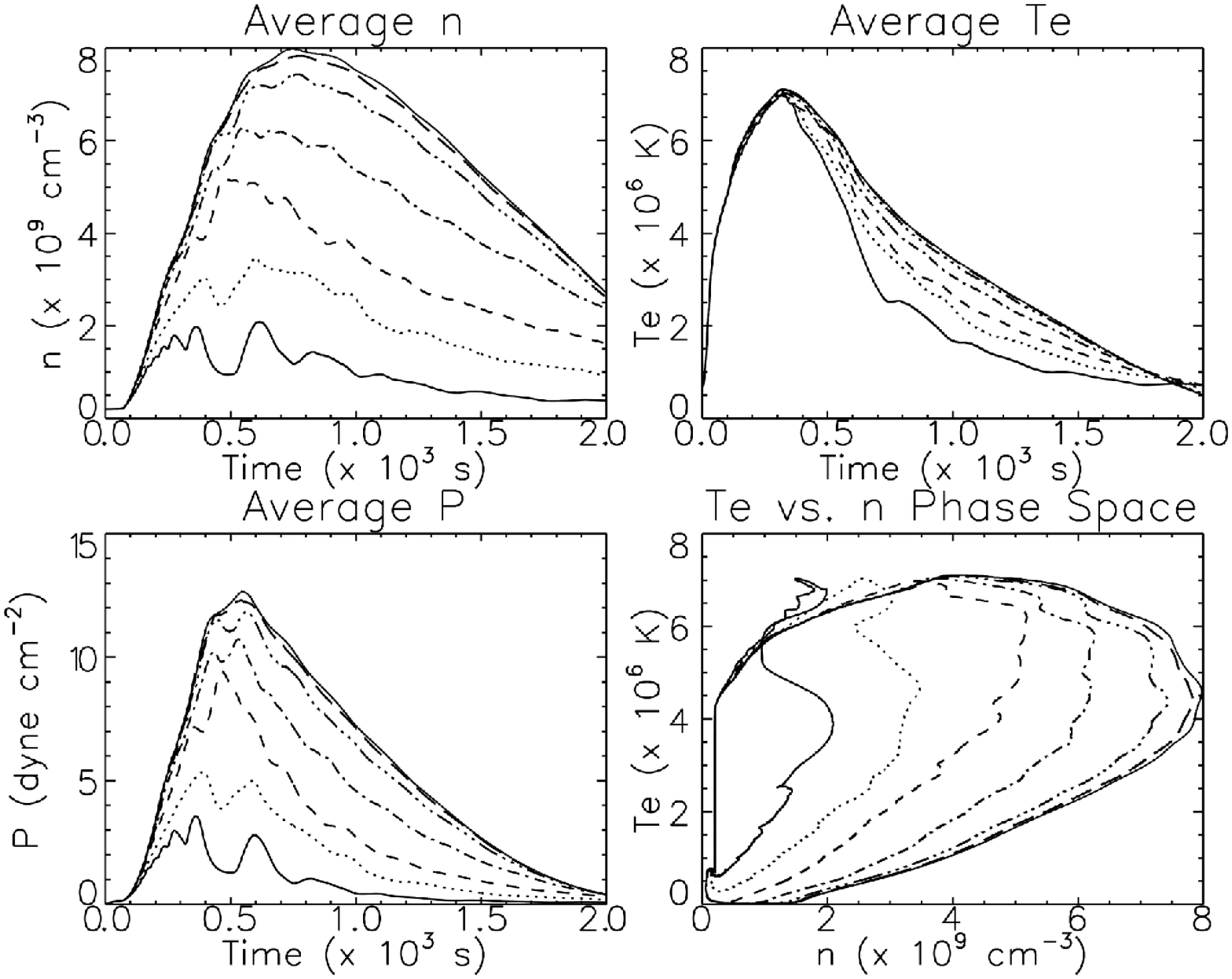}
\includegraphics[width=0.8\textwidth]{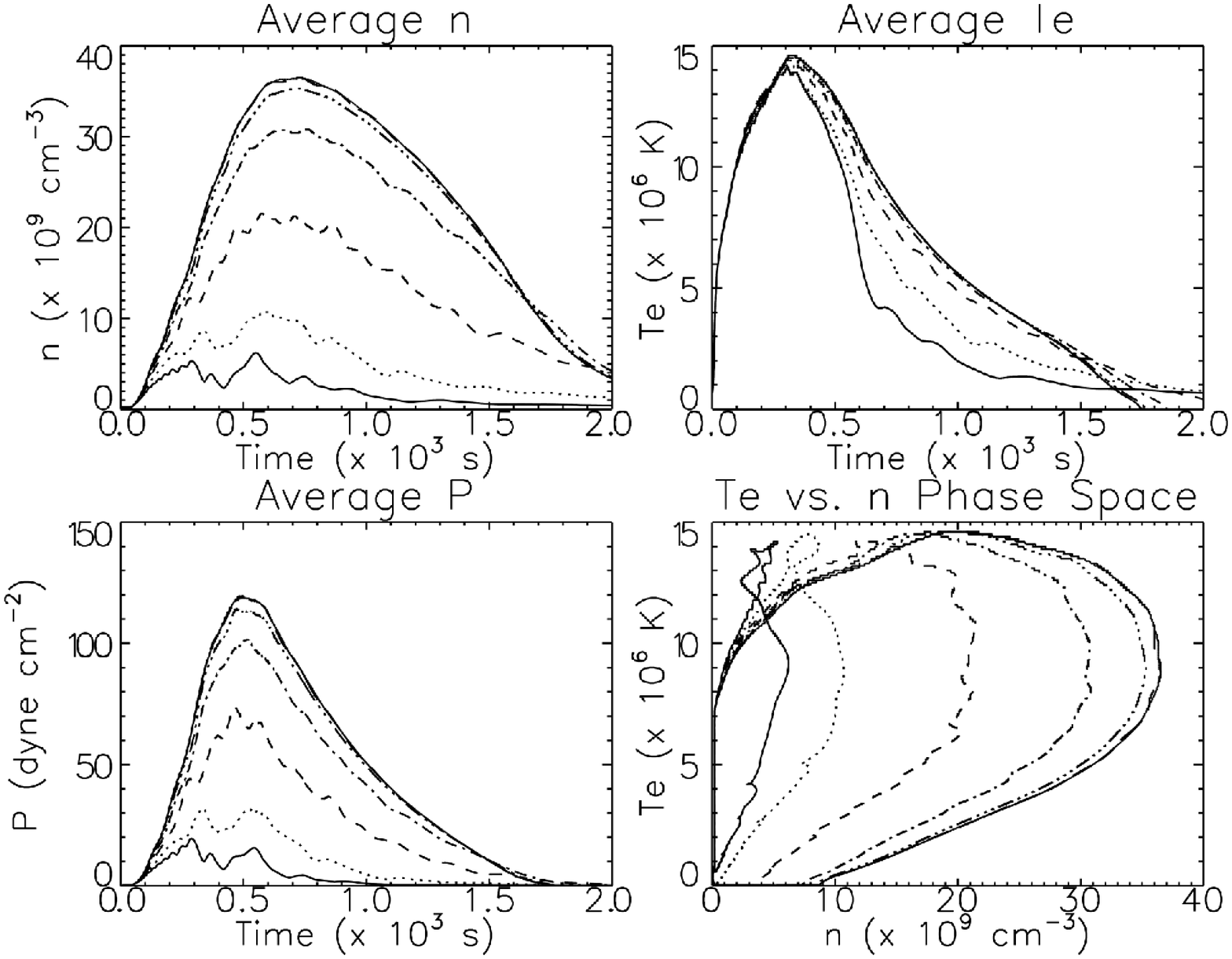}
\caption{Results for groups~5 (upper four panels) and 6 (lower four panels). Notation is the same as Figure 1}
\label{fig3}
\end{figure}

\begin{figure}
\includegraphics[width=0.8\textwidth]{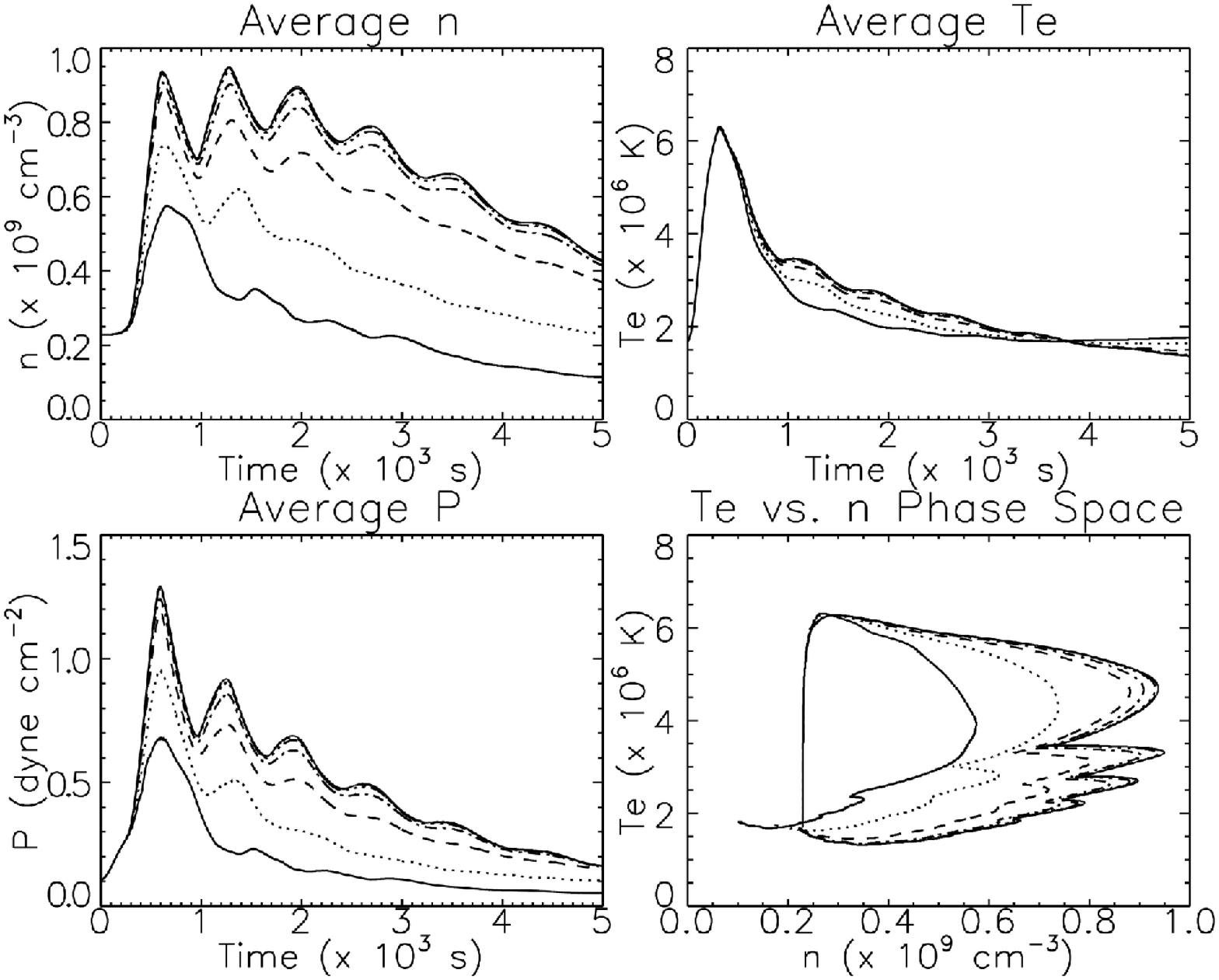}
\includegraphics[width=0.8\textwidth]{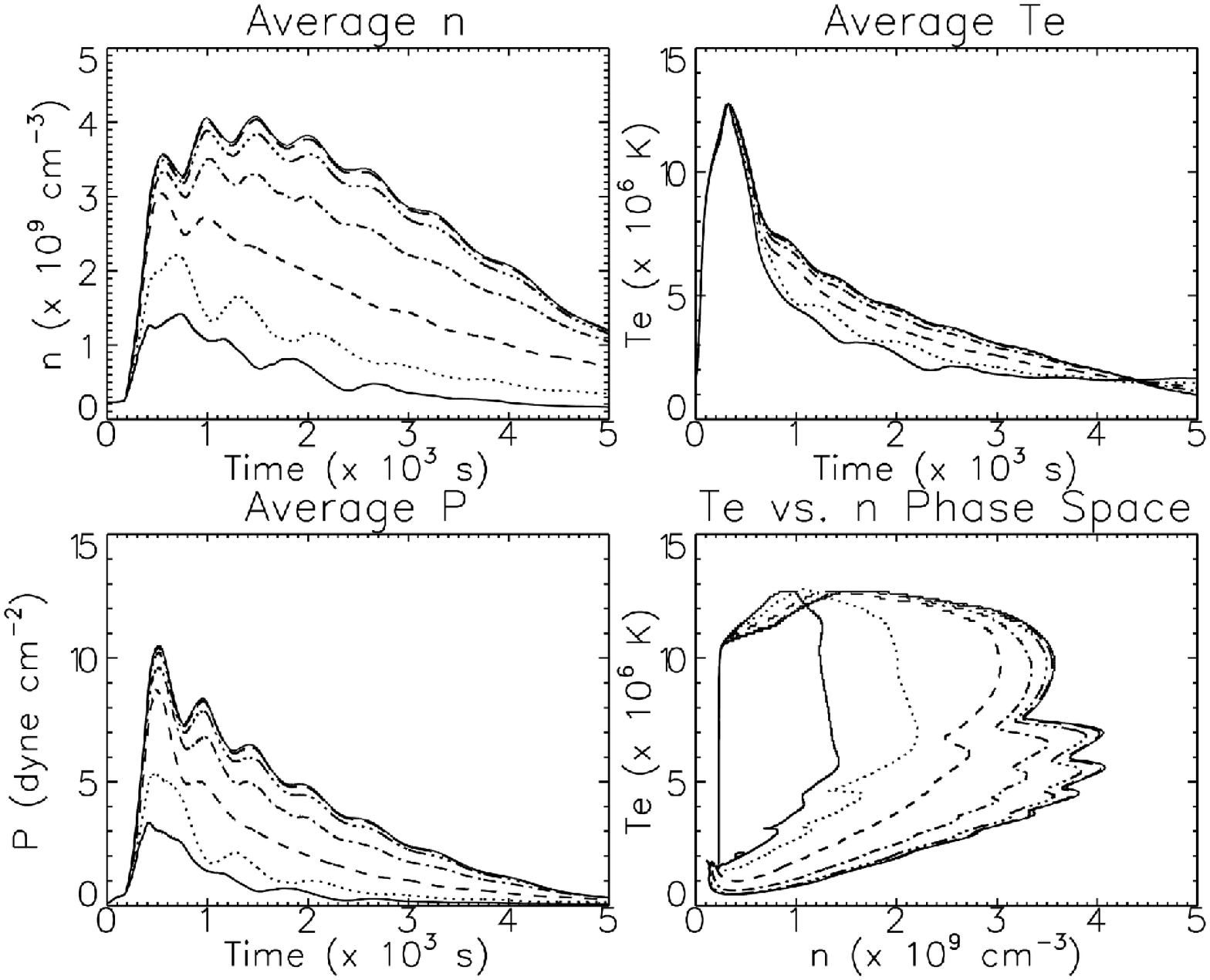}
\caption{Results for groups~11 (upper four panels) and 12 (lower four panels). Notation is the same as Figure 1}
\label{fig4}
\end{figure}

Figures~\ref{fig3} and \ref{fig4} show the moderate and extreme volumetric heating cases for loops of total length 60 (groups~5 and 6) and 180~Mm (groups~11 and 12) respectively, with the energy released over an extended period of 600~s, more akin to flaring timescales. Order of magnitude differences between the density curves for $RL=[0,12]$ can be seen for group~6. Even at refinement levels of $RL=6$ ($\approx 6$~km grid cells) there are differences of 30\% (group~5) and 20\% (group~6) between the peak densities attained and the fully converged peak densities. The peak density also shifts to earlier times by a couple of hundred seconds with changing $RL$, which is particularly noticeable for group~5. The density oscillations seen in the earlier resolved results have also disappeared for groups 5 and 6 due to the short conductive damping timescale.

The differences between the density curves for groups~11 and 12 are less severe, indicating once again that the shallower gradients associated with longer loops, for a given peak temperature, can be satisfactorily handled by coarser grids on the order of at least km resolution (cell widths of 6~km in the case of group~11 and about 2~km for group~12). The phase-space plots corresponding to groups~5 and 6, in particular, provide rather dramatic examples of the need to provide adequate spatial resolution in order to accurately follow the complete coronal heating and cooling cycle. The density oscillations have re-appeared due to the increase in the conductive damping time in the longer loops and there is a little evidence for the oscillations persisting in the unresolved long loops.

In summary, analysis of groups $1-12$ shows the following: (1) shorter loops require greater spatial resolution than longer loops for a given peak temperature; (2) the coronal density is far more strongly resolution-dependent than the temperature; (3) grid cell widths of less than 1~km are required for fully-converged solutions in the case of 60~Mm loops and peak temperatures greater than 6~MK; (4) cell widths of no more than 5~km are required for reasonably accurate solutions (densities within a few percent of converged solutions) in the case of 180~Mm loops and peak temperatures exceeding 6~MK; (5) obtaining a physically realistic evolution through the complete coronal heating and cooling cycle demands adequate spatial resolution at all times.

We can thus conclude that (i) the mechanism responsible for driving the ablation of material into the corona during heating is strongly inhibited by a lack of spatial resolution, and we will examine the physics and the numerics of this process in greater detail in Section~\ref{transport}. The phase-space plot tracks the solutions all the way back to the recovery of the initial conditions and reveals significant differences in the evolutionary paths taken by the individual runs. (ii) The under-resolved cases have rapid damping of the density oscillations. An analysis of the density and temperature in the $RL = 0$ and $RL = 12$ cases shows that there is little change in the conductive cooling times, the smaller values of $T$ and $n$ in the unresolved case offsetting each other. We discuss the reason for this rapid damping in the under-resolved cases in Section~\ref{transport}.

\subsection{Energy conservation}
\label{conservation}

The first question that needs to be addressed in understanding these results is whether the low coronal density for poorly resolved cases is caused by a lack of energy conservation due to numerical under-resolution. Figures~\ref{fig1} to \ref{fig4} show the temporal evolution of the average total pressure (electron + ion) calculated in the same way as the average density and electron temperature. The total pressure can be considered a proxy for the total thermal energy in the loop, and the amount of energy ultimately transported into the corona as the system evolves, since the kinetic energy component is in general relatively small. It is clear that the total pressure, and hence the total energy and energy transported, is strongly dependent on the spatial resolution. The total energy released is the same within each group of runs, so an increasing fraction of this energy does not appear in the corona as the spatial resolution is decreased. To take an extreme example; the lower four panels in Figure~\ref{fig3}, corresponding to group~6, show differences in the total pressure of a factor of almost 6 between $RL=0$ and $RL>8$. There are also significant differences between intermediate refinement levels. 

To assess the ability of the code to conserve energy, we check whether the following condition is satisfied:

\begin{equation}
\int_0^{2L} \frac{\partial E}{\partial t} - \left[ -E_R + E_H - \rho v g_{||} \right] ds = 0,
\label{eqn1}
\end{equation}

\noindent where $E$ is the sum of the thermal and kinetic energy, $E_R$ and $E_H$ are the volumetric radiation and heating rates, respectively, and $\rho v g_{||}$ is the rate of change of gravitational potential energy. The other terms involved in the energy equation that determine $\frac{\partial E}{\partial t}$ are flux divergencies (e.g. enthalpy and thermal conduction) which spatially integrate to zero over the entire loop due to there being no mass or conductive flux through the lower boundaries. Thus they serve only to redistribute energy within the system. Therefore, if all of the energy sources, sinks and fluxes are correctly treated and accounted for then the condition described by Equation~\ref{eqn1} will be satisfied.

We have calculated the deviation from zero of Equation~\ref{eqn1} as a percentage of the total energy stored in the transition region and corona ($T>0.02$~MK) for the most extreme cases, which are Groups~[3,6,9,12]. We chose to omit the chromospheric part of the loop from the calculation in order that the large energy content of the chromosphere could not mask any significant lack of conservation in the overlying atmosphere. Between $t=0$~s and the time of peak density for each group of runs (as seen in Figures~\ref{fig1} to \ref{fig4}) we found the maximum percentage deviations from perfect conservation to be: Group~3, 0.36\%~($RL=0$), 0.20\%~($RL=12$); Group~6, 0.69\%~($RL=0$), 0.09\%~($RL=12$); Group~9, 0.05\%~($RL=0$), 0.12\%~($RL=12$); Group~12, 0.31\%~($RL=0$), 0.26\%~($RL=12$). In addition, the temperature vs. density phase-space diagrams in Figures~\ref{fig1} to \ref{fig4} show that following the conclusion of the impulsive heating, when only the background heating used to find the initial hydrostatic conditions remains, the code does a good job of recovering the initial conditions by returning the solution to the starting point in the phase space. Thereafter, the atmosphere is held in hydrostatic equilibrium with only residual background flows having a Mach number $M<<1$.

We find that energy is conserved to better than 1\%, even for $RL=0$, and so the large discrepancies in the total coronal thermal energy indicated by Figures~\ref{fig1} to \ref{fig4} are not due to a lack of spatial resolution directly (because the code is evidently doing a good job of accounting for all energy sources, sinks and fluxes), but the lack of spatial resolution may be driving the solutions into different physical regimes (e.g. strong vs. weak ablation), as we now discuss. Whether energy equations implemented in 3D codes that do not use the conservative form do so well is unclear.

\subsection{Energy transport by enthalpy and thermal heat flux}
\label{transport}

\begin{figure}
\includegraphics[width=0.8\textwidth]{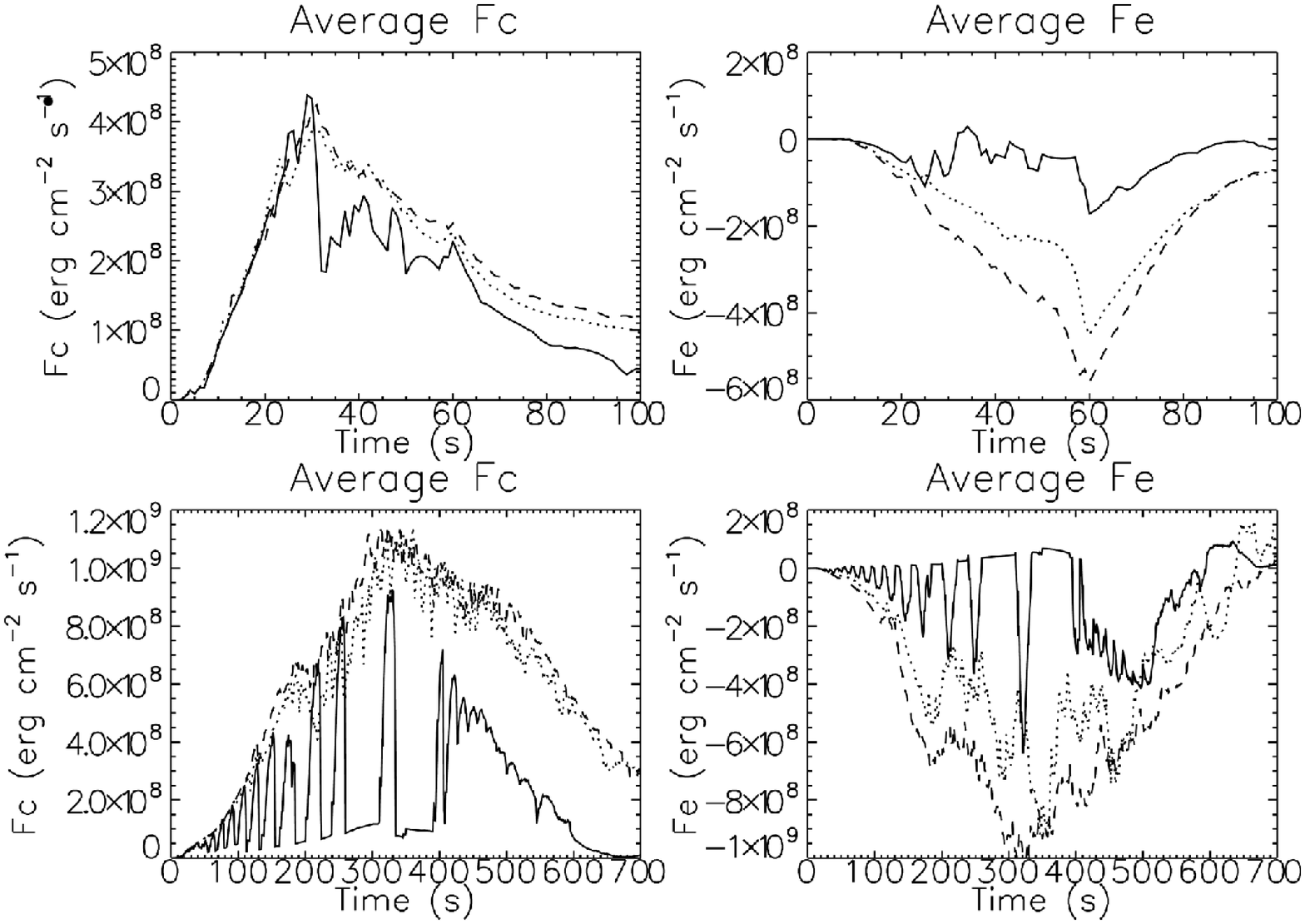}
\includegraphics[width=0.8\textwidth]{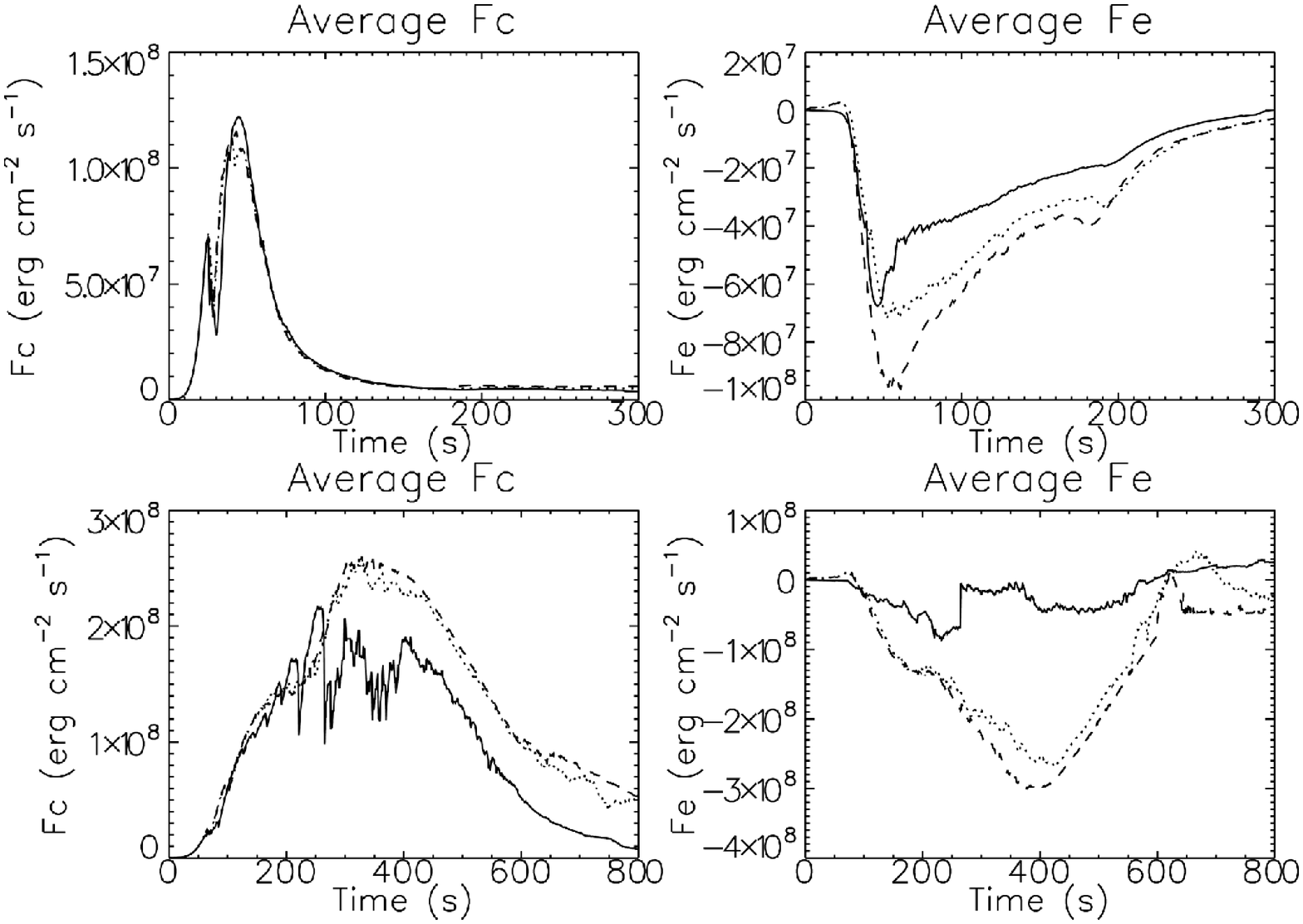}
\caption{Results for groups~3 (upper row), 6 (second row), 9 (third row) and 12 (lower row). The left-hand column shows the average heat fluxes and the right-hand column the average enthalpy fluxes in the transition region. The curves for $RL=0$ (solid), $RL=6$ (dotted) and $RL=12$ (dashed) are plotted.}
\label{fig5}
\end{figure}

We now turn to a consideration of the different physical regimes that may dominate the numerical solutions for the range of $RL$ studied here. Since the major differences between these runs are a consequence of the transition region resolution then it makes sense to focus our investigation on this part of the domain. Furthermore, we demonstrated in our earlier work \citep{Bradshaw_2010a} that it is the demands of the transition region which dominate the energetics and the dynamics of the loop during all phases of its evolution. Figure~\ref{fig5} shows the average heat flux (left-hand column) and the average enthalpy flux (right-hand column) through the transition region for groups~3, 6, 9 and 12. They were calculated by spatially integrating the heat flux from the location of the transition region~/~corona interface down to the foot-point (the first grid cell in which the temperature reaches or falls below 20,000~K) and then dividing by the transition region width. The location of the transition region~/~corona interface can be found in a reasonably straightforward way during this period by identifying the grid cell in which the divergence of the heat flux transitions from positive (corona) to negative (transition region), indicating that energy is being removed from the heat flux to power the transition region radiation. These average quantities were calculated for the right-hand leg of the loop and so positive fluxes are down-flowing (toward the foot-point) and negative fluxes are upflowing (toward the apex). Their temporal evolution is shown for the entire heating phase and part of the ablative phase (coronal cooling by thermal conduction). At later times, when the ablated material from opposite legs of the loop collides at the apex and triggers acoustic waves (the density oscillations already noted, with accompanying localised temperature perturbations) then this interface becomes far more difficult to locate.

An examination of the left-hand column of Figure~\ref{fig5} shows that the ability of the transition region to support a substantial heat flux is almost independent of the spatial resolution, particularly in the case of the groups of runs for the longer loops (groups~9 and 12) where the gradients are shallower. Recall that these results are for the most extreme heating scenarios chosen for our study. Even in the case of short loops (groups~3 and 6) the heat fluxes for the $RL=[6,12]$ runs are in very close agreement, given the orders of magnitude difference in the spatial resolution (a factor of 64). Once again, the discrepancies between the coronal density and the total energy for different values of $RL$ are not due to any significant heat flux limiting caused by inadequate spatial resolution in the transition region.

The right-hand column of Figure~\ref{fig5} reveals a clue to what is actually behind the discrepancies. The common assumption is that the heat flux drives {\it chromospheric} ablation, where the excess energy carried by the heat flux that is not radiated by the transition region heats the chromosphere (a poor radiator), which then expands upwards and carries with it an enthalpy flux in the opposite direction to the heat flux. However, the return enthalpy flux is severely diminished at low transition region spatial resolution, which is indicative of an alternative scenario. If it were purely chromospheric ablation then, since most of the heat flux is accounted for even at small $RL$, the resulting ablation and enthalpy flux should reflect the agreement between the heat fluxes for different values of $RL$. Evidently this is not the case and a significant amount of energy is somehow lost before it is converted into enthalpy at low spatial resolution.

The true physics of ablation are as follows. At low spatial resolution, where the transition region is effectively a discontinuity, the heat flux jumps from the corona to the chromosphere where the incoming energy is strongly radiated (the chromosphere is an inefficient radiator relative to its density, but it {\it is} dense), leaving little left over to heat the plasma and drive ablation. At high spatial resolution the heat flux passes through the transition region in a series of steps and at each step some energy is radiated, some goes into local heating and expansion~/~ablation, and the remainder continues on. More and more energy is radiated with increasing depth in the transition region since the volumetric loss rate scales as $n^2$. Ultimately, depending on the magnitude of the initial heat flux, a proportion of the original heat flux will make it to the chromosphere, but a significant contribution to the upward material and enthalpy flux will have been made by transition region plasma. This contribution is neglected in low resolution runs, with the resultant discrepancies between the coronal density and the total energy. It may also partially explain why observational evidence of the upflows that must somehow fill the corona has been so difficult to provide; there is no single, dominant source of plasma and instead the corona is filled in a more piece-meal fashion.

\begin{figure}
\includegraphics[width=1.0\textwidth]{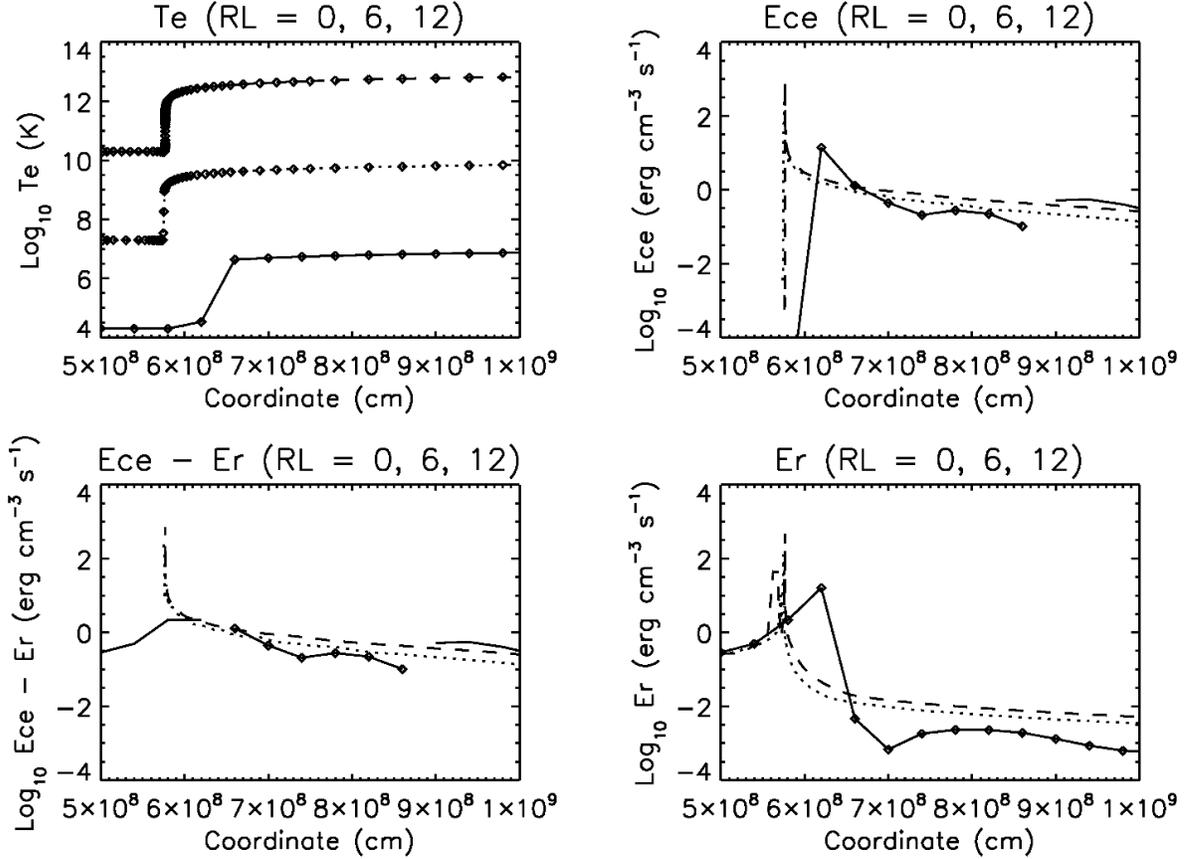}
\caption{Results for group~3 at $t=50$~s, when the coronal density is approximately half of its peak value. Upper-left panel: electron temperature profile in the foot-point region for $RL=0$ (solid), $RL=6$ (dotted) and $RL=12$ (dashed); the curves of the stack-plot are offset by 3.0~dex ($RL=6$) and 6.0~dex ($RL=12$); the spacing of the diamond symbols indicates the spatial resolution. Upper-right panel: divergence of the heat flux. Lower-right panel: radiative losses. Lower-left panel: estimate of the net heating; where this quantity is greater than zero then energy is available to drive an upflow; for $RL=0$ the solid line connected by diamond symbols shows where the net heating is greater than zero and the solid line without diamonds shows where it is less than zero.}
\label{fig6}
\end{figure}

Figure~\ref{fig6} confirms this conclusion. The upper-left panel shows the electron temperature profile in the foot-point region for the $RL=[0,6,12]$ runs of Group~3, approximately halfway through the ablation phase. Group~3 has been selected for detailed analysis because it imposes very severe constraints on the spatial resolution (see Table~\ref{table1} and Figure~\ref{fig1}). The temperature gradient is extremely poorly resolved in the $RL=0$ case to the extent that there are no data points for temperatures of O(10$^5$)~K. The upper-right panel shows that for $RL=0$ a large amount of energy is deposited by the incoming heat flux in the grid cell at $s=6.2\times10^8$~cm, corresponding to $T=10^{4.5}$~K, and then it steeply falls away. The lower-right panel shows strong radiation from the same grid cell and the lower-left panel shows that at this critical grid cell there is no energy left over to drive ablation. For $RL=[6,12]$ it is clear that there is a net heating throughout the region of the steep temperature gradient and so energy is available to ablate the increasingly dense plasma found there, driving it upward and leading to enhanced coronal densities compared with $RL=0$. The transition region~/~corona interface is properly described as the location where the heat flux transitions from a sink (the corona) to a source (the transition region) as defined by \cite{Bradshaw_2010a}. The upper-right panel of Figure~\ref{fig6} shows that this extends beyond $s=10^9$~cm for the $RL=[6,12]$ runs (actually to $1.5\times10^9$~cm) and to about $9\times10^8$~cm for the $RL=0$ run. The steep temperature gradient may then be thought of as the lower transition region and, since the energy deposited by the heat flux in the upper transition region is quite similar for $RL=[0,6,12]$, then the dominant source of coronal plasma is the lower transition region. It is also interesting to note that under-resolving the transition region evidently leads to an underestimate of its spatial scale, despite the similarities in the time evolution of the average temperature between the runs at this time. Other groups with severe resolution constraints show the same result. Thus the mechanics of ablation are far more dependent on resolving the lower transition region and treating its radiative properties correctly than being overly concerned with the energetics of the chromosphere. This is the underlying reason why relatively simple treatments of the chromosphere, particularly in flare modeling, can yield similar results to those that are more sophisticated; the chromosphere does not make the dominant contribution. Of course, the more sophisticated treatments are necessary for detailed spectral modeling of chromospheric emission.

The nature of the interaction between the corona and the chromosphere, mediated by the transition region, also explains why the density oscillations, caused by ablated material sloshing back and forth in the corona, are rapidly damped in the under-resolved cases. When there is inadequate spatial resolution the wave energy falls over the transition region cliff to be radiated away by the chromosphere, whereas in the well-resolved cases the effective refractive index of the transition region is greater, leading to more of the wave energy being reflected from the steep density gradient and back into the corona.

\section{Discussion}
\label{discussion}

The results presented here demonstrate that the principal effect of lack of transition region resolution is on the coronal density. In this Section we explore a number of questions that arise and are of importance to constructing numerical models of the outer solar atmosphere.

\subsection{Implications for multi-dimensional MHD models}

We have demonstrated in this paper that one can run 1D HD models using highly refined adaptive grids (tens of m) for several hours real (solar) time. The situation with multi-dimensional MHD models is more complicated for several reasons: 

(i) A limit on resolution is imposed by the scale of the simulation being run. A contemporary massively parallel code such as Bifrost \citep{Gudiksen_2011} can run something of the order $512^3$ cells (depending on available resources) and the grid can be stretched in one dimension, permitting a fixed concentration of points where any transition region is expected. [Adaptive grids are more complex to implement in 3D MHD since there are many places that one might require high resolution, such as current sheets, in addition to the transition region.] As examples of contemporary 3D codes, \cite{Gudiksen_2005a} have cells of 150~km in the vertical direction, \cite{Bingert_2011} 230~km, decreased subsequently by \cite{Zacharias_2011} to 75~km. These correspond to $RL = 2 - 4$. All these papers discuss relatively cool loops, with temperatures of order 1 MK, so are closest to our group 1 (weakest heating for short loops) where we find $RL = 4$ is needed (Table~\ref{table1}). This suggests that the resolution is adequate for these particular cases, though even marginally higher temperatures will result in inadequate resolution. The Bifrost code solves for thermal conduction on a sub-grid with the potential for good resolution equivalent to $RL=12$, however at the present time a resolution of up to
25~km is used (Gudiksen, private communication, 2013): we discuss other aspects of their implementation technique below.

(ii) Other solutions have been proposed in order to sidestep the need for high resolution, such as  an artificial reduction of $\kappa_0$. This would appear to permit significantly greater timesteps, but by mitigating the cooling role of conduction, higher coronal temperatures are maintained, which will tend to push the timestep back to shorter values. \cite{Gudiksen_2011} also propose lowering the thermal conduction coefficient, but comment that ``the results must be carefully analysed''. \cite{Linker_2001} proposed an alternative formalism for $\kappa (T)$:

\begin{equation}
\kappa(T) = \kappa_0 \left [ s T^{5/2} + (1 - s) T^\alpha T_{\mbox{mod}}^{5/2 - \alpha} \right ]
\end{equation}
\noindent
where
\begin{equation}
s(T) = \frac{1}{2} \left [ 1 + \tanh \left ( \frac{T - T_{\mbox{mod}}}{\Delta T_{\mbox{mod}}} \right ) \right ]
\end{equation}

\noindent Here $T_{\mbox{mod}}$ is a typical transition region temperature (0.2~MK say), and above (below) this temperature the heat flux approaches the Spitzer value (is decreased). For typical active region temperatures, solution of the static equilibrium loop equations leads to a decrease of order two in the coronal density, as also noted by \cite{Linker_2001}. Subsequently, \cite{Lionello_2009} argues that if the radiative losses are also diminished such that $\kappa(T) R_L(T)$ is unchanged, one can recover proper coronal densities: indeed for static loops, we find the difference is $< 1 \%$. However, it is unclear how successfully this formalism translates into dynamic evolution. The increase in coronal density after heating arises precisely because the transition region is not in equilibrium: the downward heat flux exceeding the radiative losses, leading to an upward enthalpy flux \citep[see][]{Klimchuk_2008,Cargill_2012}. In the decay of impulsive events, the coronal properties are governed by a downward enthalpy flux responsible for the transition region radiation: conduction plays a minimal role \citep{Bradshaw_2008,Bradshaw_2010a,Bradshaw_2010b}.

In addition, the physical definition of the transition region is such that the transition region~/~corona interface is located where the supporting flux transitions between a sink and a source \citep{Bradshaw_2010a,Bradshaw_2010b}. The transition region is {\it not} simply anywhere that $T<1$~MK. The characteristic temperature of the transition region can change dramatically during the heating and cooling cycle; from several MK in hot AR cores and during flares, to below 1~MK during the later stages of cooling. It is therefore unclear how one should choose the quantity $T_{\mbox{mod}}$ at each stage of the numerical calculation. Studies are clearly required to validate this approximation: its benefits are considerable in terms of the run time of a 3D code and even a 1D code (e.g. for flares); its disadvatages are that any transition region spectral line properties are unlikely to be relevant.

(iii) Another issue that may arise with 3D MHD codes is the reluctance to use the conservative form of the energy equation. This arises because in a conservative MHD code, one calculates the plasma pressure by subtracting the magnetic field energy density (and kinetic energy density) from the total energy density: for a low $\beta$ plasma, typical of the corona, the pressure is then the difference between two large numbers, leading to a possibility of negative pressure. There are in fact forms of the energy equation that get around this \citep[see][]{Cargill_2000}, but these run into difficulties in modeling shocks. In under-resolved non-conservative simulations, it is unclear whether the lack of conservation of energy is a problem. Some benchmarking is required.

(iv) Some 3D codes use operator splitting, or time-splitting, in the interests of computational efficiency (e.g. not solving the MHD equations on a conductive timescale). Here the various parts of the energy equation are solved separately. For example, \cite{Tam_2013} solve the conduction and radiation parts of the equation separately, and then feed the new temperature back into the MHD equations. \cite{Gudiksen_2011} solve the conduction equation implicitly on a fine grid, and then generate a smoothed solution to feed into the MHD equations. It is unclear to us whether such procedures guarantee, for example, the correct interaction between a downward conductive flux, and the response through an upflow. For the benign problems discussed in \cite{Bingert_2011} and \cite{Gudiksen_2011} there may not be an issue. But a benchmarking of such procedures seems to be needed.

To close this part of the discussion, we recommend that those using 3D MHD codes to study the interaction between corona, transition region and chromosphere carry out a set of ``resolution tests'' as we have done here with more extreme coronal temperatures than those published. This can be accomplished by reducing the 3D code to a 1D one, while keeping the overall 3D grid structure, with the remaining coordinate aligned along a straight magnetic field. Such a comparison will provide confidence in how the 3D codes address this important aspect of modeling the solar atmosphere.

\subsection{Implicit solvers: a 3D panacea?}

Our numerical experiments have employed a fully explicit 1D HD code, which forces us to resolve the principle physical timescales of the system at each stage of its evolution. Other workers have resorted to implicit solvers for their 3D MHD codes \citep[e.g.][]{Botha_2011,Gudiksen_2011,Tam_2013}. The advantage of, for example, an implicit Crank-Nicolson scheme is that it is guaranteed to be numerically stable. Some sort of iteration is undertaken in multi-dimensions \citep[e.g. a Successive Over-Relaxation (SOR) scheme:][]{Botha_2011}, with a criteria for convergence being a small discrepency in the temperature (either averaged, or pointwise) between iterations. This can slow down an implicit scheme if the number of iterations is large, or, worse, if convergence never occurs.

Such implicit schemes are useful for a slowly-evolving system  or one seeking equilibria, where long wavelength behaviour dominates, and one is not concerned with temperature fluctuations over one or two grid points. However for general applications one must always bear in mind the various physical timescales that are of interest. The most important one here is the time it takes a conduction front to cross a grid cell, where the cell is narrow enough to resolve any temperature gradient. For a dynamic situation, one needs a ``Courant condition'' for a conduction front: $\Delta t(\mbox{implicit}) < \mbox{min}(\Delta s /  V_c)$ where $V_c$ is the velocity of the conduction front. For an analytic solution such as the Zel'dovich thermal wave $V_c$ is known. However, in a simulation, one needs to calculate both $\Delta s$ and $V_c$ in real time. The former is accomplished through an adaptive grid. The latter is more challenging. An alternative way to impose this condition would be to require that the change in temperature in any cell be limited to some reasonably small fraction (e.g. 20\%), and adjust the timestep accordingly. Such a secondary iteration (in addition to the one needed to solve the diffusion equation itself) would need careful implementation to avoid crippling computational costs.

\subsection{Interpretation of data}

Besides the obvious implications for calculating the response of the lower layers of the atmosphere to heating and the subsequent energy transport, these results are collectively significant for spectral modeling and emission measure calculations. The differences between the electron temperature curves, though small, may lead to very different predictions for ionisation states and spectral line intensities because the temperature-dependent contribution functions tend to be very sharply peaked. Consequently, small temperature differences matter. The much greater differences between the density curves are even more problematical. The emission measure and spectral line intensity scales as $n^2$, so differences are non-linearly amplified, and time-dependent calculations of the ionisation state are likely to be strongly affected due to the density dependence of the collisional ionisation~/~recombination timescales. It is not difficult to identify the potential for forward modeling studies, in which numerical predictions are compared with real observational data, to yield misinterpretations and erroneous conclusions.

\section{Summary and conclusions}
\label{summary}

This paper has addressed the role of numerical resolution in models of impulsive heating in the solar corona. A range of event sizes have been discussed, ranging from nanoflares to microflares to small flares. The principle effect of lack of numerical resolution is on the coronal density, which is decreased significantly in unresolved cases. The temperature also decreases, though not by as much. The implications are very significant for forward modeling, with the intensity of spectral lines being underestimated. Also, the deviations of the temperature will lead to the incorrect prediction of dominant lines due to the sharply peaked contribution functions.

With a proper adaptive grid, we believe that the resolution problem is probably managable today for 1D HD models with a well structured code, and will certainly be managable in the future. For higher-dimension MHD codes, present day models deal with a relatively cool corona (1~MK), and we noted earlier that adequate resolution was probably achieved with a structured (as opposed to adaptive) grid. However, the required resolution scales as a very high power of temperature, leading to (a) structured grids ultimately becoming uneconomical and (b) the need for adaption is equally pressing with other areas of the physics, such as in current sheet, leading to a huge accounting problem. Some potential solutions to this have been discussed above, including modifying the conduction coefficient and sub-cycling the conductive timestep.

Finally, we urge those running 1D HD and especially 3D MHD simulations to undertake a series of tests as we have done here to establish what resolution is really needed. 

\acknowledgements

SJB acknowledges the support of NASA through the Supporting Research and Technology (SR\&T) program. PC thanks Alan Hood for a number of useful discussions. The early ideas for this work arose from discussions that took place during an international team meeting (led by SJB and Dr. Helen Mason) at the International Space Science Institute (ISSI) in Bern. We thank ISSI for their support and our team members for their contributions. We also thank the referee for helpful suggestions that improved the presentation.

\appendix
\section{The HYDrodynamics and RADiation (HYDRAD) code}
\label{AppA}

\subsection{The hydrodynamic equations}

The evolution of the HYDRAD code used in this paper has been described in various publications \citep{Bradshaw_2003a,Bradshaw_2003b,Bradshaw_2006,Bradshaw_2011b}. Given its now widespread use, it is useful to describe all the developments of HYDRAD and the current capabilities of the code in one location. HYDRAD solves the one-dimensional, multi-fluid hydrodynamic equations for a full loop of arbitrary geometry. The mass and momentum equations reduce to the following conservative form:  

\begin{equation}
\frac{\partial \rho}{\partial t} + \frac{\partial}{\partial s}\left(\rho v\right) = 0,
\label{eqnA1}
\end{equation}

\begin{equation}
\frac{\partial}{\partial t}\left(\rho v\right) + \frac{\partial}{\partial s}\left(\rho v^2\right) = -\frac{\partial}{\partial s}\left(P_e + P_i\right) + \frac{\partial}{\partial s}\left(\frac{4}{3} mu_i \frac{\partial v}{\partial s}\right) - \rho g_{||},
\label{eqnA2}
\end{equation}

\noindent where current and charge neutrality have been imposed, $t$ and $s$ are time and the field-aligned spatial coordinate, $\rho$ is the ion mass density ($\rho = m_i n$: $m_e \ll m_i$), $v$ is the bulk velocity of the flow, and $g_{||}$ is the field-aligned gravitational acceleration. Here $m_i$ is the average ion mass, and accounts for the fact that while hydrogen is by far the most abundant element the presence of heavy metals has a non-negligible influence on the mass density and the bulk flow. Ion viscosity is included using the classical Spitzer viscosity coefficient. Electron viscosity is negligible. There are separate energy equations for the electrons (subscipt `e`) and ions (subscript `i`):

\begin{equation}
\frac{\partial E_e}{\partial t} + \frac{\partial}{\partial s}\left[\left(E_e + P_e\right)v\right] = -\frac{\partial F_{ce}}{\partial s} + v\frac{\partial P_e}{\partial s} + \frac{k_B n}{\gamma-1}\nu_{ie}\left(T_i-T_e\right) - R + H,
\label{eqnA3}
\end{equation}

\begin{equation}
\frac{\partial E_i}{\partial t} + \frac{\partial}{\partial s}\left[\left(E_i + P_i\right)v\right] = -\frac{\partial F_{ci}}{\partial s} - v\frac{\partial P_e}{\partial s} + \frac{k_B n}{\gamma-1}\nu_{ie}\left(T_e-T_i\right) + \frac{\partial}{\partial s}\left(\frac{4}{3} mu_i v \frac{\partial v}{\partial s}\right) + \rho v g_{||}.
\label{eqnA4}
\end{equation}

\noindent where:

\begin{equation}
E_e = \frac{P_e}{\gamma-1}, ~~E_i = \frac{P_i}{\gamma-1} + \frac{1}{2}\rho v^2.
\label{eqnA5}
\end{equation}

\noindent and closure is imposed by:

\begin{equation}
P_e = k_B n T_e,  P_i = k_B n T_i, ~~~ 
\label{eqnA6}
\end{equation}

\noindent where $T_i$ and $T_e$ are the ion and electron temperatures, and $k_B$ is Boltzmann's constant. The first term on the right-hand side of equations~\ref{eqnA3} and \ref{eqnA4} represents the thermally conducted flux carried by each particle species. The electron flux is written as:

\begin{equation}
F_{ce} = - \kappa_{0e} T_e^{5/2} \frac{\partial T_e}{\partial s}
\label{eqnA7}
\end{equation}

\noindent and the ion flux in an analagous form. The Spitzer coefficient for the electron thermal conduction is $\kappa_{0e} = 7.8\times10^{-7}$ and we assume that protons dominate the ion thermal conduction, giving $\kappa_{0i} = 3.2\times10^{-8}$. A flux-limiter is built into the numerical code to ensure that the free-streaming limit is not exceeded and is discussed below \citep[see also][]{Patsourakos_2005,Bradshaw_2006}. The second term on the right-hand side of equations~\ref{eqnA3} and \ref{eqnA4} $P_e$, represents the electric potential; the energy given-to~/~removed-from the particle species by the electric field when the separation between them becomes too large~/~small.  The third term represents the thermal equilibration between the particle species via collisions, where $\nu_{ie}$ is the Coulomb collision frequency:

\begin{equation}
\nu_{ie} = \frac{16\sqrt{\pi}}{3} \frac{e^4}{m_e m_i} \left(\frac{2k_B T_e}{m_e}\right)^{-\frac{3}{2}} n \left(\ln \Lambda\right),
\label{eqnA8}
\end{equation}

\noindent where $\ln \Lambda$ is the Coulomb logarithm. The gravitational potential energy has been included for the ions and neglected for the electrons. $R$ is the optically-thin radiative emission, which can be calculated using a set of piece-wise power laws, a fit to radiative loss curves calculated by the Chianti atomic database for equilibrium ionization or a full non-equilibrium treatment of the radiating ions (thermal bremmstrahlung is also included in the calculation). $R$ can also be calculated for optically-thick conditions in the lower atmosphere using the treatment given by \cite{Carlsson_2012} and in this case the presence of neutral species is accounted for (e.g. number density of neutrals, ions and electrons, inter-species collisions, thermal conduction of neutral hydrogen). However, in the present work the chromosphere is maintained at a uniform temperature (20,000~K) by reducing the optically-thin radiative losses to zero over a specified temperature interval (100~K) above the chromospheric temperature \citep{Klimchuk_1987}. $H$ is a parameterized, volumetric heating function, that depends on the field-aligned coordinate and~/~or time.

\subsection{Finite difference schemes}

The transport terms of the hydrodynamic equations are calculated using Barton's method for monotonic transport and a staggered leapfrog algorithm to perform the time integration.
\cite{Centrella_1984} provide a detailed discussion of Barton's method, which is a simpler and more rugged version of the method devised by \cite{van_Leer_1977}, and which works well even on highly discontinuous hydrodynamic problems. For a uniform grid, the method is 2nd order in space away from the discontinuities and 2nd order in time. At discontinuities it reduces to first order in space to maintain fidelity to the nature of the solution in that region. All other ordinary and source~/~sink terms are spatially differenced and time integrated to 2nd order using standard Runge-Kutta methods \citep{Press_1992}.

For a uniform grid, the conductive flux $F_c$ is calculated at the edge of cell $j$ by adopting the following discretization:

\begin{equation}
F_{c,j-1/2} = - \kappa_0 T_{j-1/2}^{5/2} \frac{T_j - T_{j-1}}{\Delta s_j},
\label{eqnA9}
\end{equation}

\noindent where:

\begin{equation}
T_{j-1/2}^{5/2} = \left[ \left( T_j + T_{j-1} \right) / 2 \right]^{5/2}.
\label{eqnA10}
\end{equation}

\noindent The volumetric rate of energy change in the grid cell due to thermal conduction is then given by:

\begin{equation}
- \frac{\Delta F_{c,j}}{\Delta s_j} = - \frac{F_{c,j+1/2} - F_{c,j-1/2}}{\Delta s_j}.
\label{eqnA11}
\end{equation}

For a non-uniform grid, the conductive flux $F_c$ is calculated at the edge of cell $j$ in such a way as to limit the potential loss of spatial order in the derivatives that is known to arise when there are variations in the widths of adjacent grid cells \citep{Oran_2000}. The temperature $T_{j-1/2}$ is first found by interpolation using a four-point stencil ($j-2$, $j-1$, $j$, $j+1$); however, if this 3rd order interpolation introduces a new local maximum or minimum then $T_{j-1/2}$ is calculated using a linear interpolation ($j-1$, $j$). Using the same four-point stencil the temperature $T^*$ is found at the position $s_j - \Delta s_j$ (again, reduced to a linear interpolation if a new local max. or min. is introduced). The heat flux at the cell boundary is then given by:

\begin{equation}
F_{c,j-1/2} = - \kappa_0 T_{j-1/2}^{5/2} \frac{T_j - T^*}{\Delta s_j}.
\label{eqnA14}
\end{equation}

\noindent This procedure is effectively equivalent to interpolating the temperature onto a locally uniform grid, centered at $j-1/2$, where the cell width is $\Delta s_j$. Note that this procedure is used by HYDRAD to calculate all spatial gradients (except for the transport terms) on a non-uniform grid (e.g. $dP/ds$, etc.).

\subsection{The adaptive grid}

An important aspect of HYDRAD is its adaptive grid, which is flexible enough to increase the numerical resolution wherever needed and so it can resolve as many discontinuities and~/~or steep gradients, for example, as arise in the calculation at any time. Restriction (refinement) is carried out according to user-defined tolerances (e.g. greater than 10\%) on cell-to-cell differences between some or all (again, user-definable) of the quantities $\rho$, $E_e$ and $E_i$. When HYDRAD detects that a grid cell needs to be restricted then it is split into two. The quantities $\rho$, $\rho v$, $E_e$ and $E_i$ are calculated in the left-most of the two new cells by interpolation from the original cells $j-2$, $j-1$, $j$ and $j+1$. Similarly, these same quantities are calculated in the right-most of the two new cells by interpolation from the original cells $j-1$, $j$, $j+1$ and $j+2$. Cell $j$ is then removed from the linked-list that connects the grid cells and the two new cells are inserted in its place. The two new cells are also tagged with a unique identifier for each refinement level so that only the same two cells that resulted from the splitting of a single cell can be merged to form a single cell under prolongation. The user can specify the number of times that a grid cell can be split and this is the maximum refinement level. When HYDRAD detects that two grid cells ($j$ and $j+1$) with the same unique identifier satisfy the criteria to be merged into a single cell (e.g. cell-to-cell differences between quantities falls to less than 5\%) then the prolongation operator calculates $\rho$, $\rho v$, $E_e$ and $E_i$ in the new cell by interpolation from the original cells $j-1$, $j$, $j+1$ and $j+2$. Cells $j$ and $j+1$ are then deleted from the linked-list and the new cell inserted in their place. The user can select linear interpolation for the restriction and prolongation operators if they expect that sharp features such as shocks and~/~or steep gradients might arise in their solutions. In this case, just the inner two points of the stencils described above are used. The four-point polynomial interpolation routine also checks for new local minima or maxima and reduces to a linear interpolation if necessary. The adaptive grid algorithm employed by HYDRAD limits the variation in the widths of adjacent cells to a single refinement level as a precaution against the loss of order in calculating spatial derivatives on non-uniform grids \citep{Oran_2000}.

\subsection{Alternative finite difference schemes for the heat flux}

While Equations~\ref{eqnA9} and \ref{eqnA10} are a conventional way to finite difference the conduction term, they are not unique. One option is to write Equation~\ref{eqnA10} as:

\begin{equation}
T_{j-1/2}^{5/2} = \left[ \left( T_j^{5/2} + T_{j-1}^{5/2} \right) / 2 \right]
\label{eqnA12}
\end{equation}

\noindent A second is to define the variable $u = T^{7/2}$ so that the conduction term is:

\begin{equation}
F_c = - \frac{2}{7} \kappa_0 \frac{\partial^2 u}{\partial s^2}
\label{eqnA13}
\end{equation}

\noindent and finite difference this in the usual 3-point manner. We have compared these with Equations~\ref{eqnA9} and \ref{eqnA10} by solving the Zel'dovich problem of a propagating heat front \citep{Zeldovich_1967, Reale_1995}. For a resolved front, the solution agrees well with the analytical one for all three methods. For an unresolved heat front (i.e. a case where the major temperature jump at the leading edge of the front is over one cell), the three finite difference schemes give heat fluxes at the leading edge that differ by up to a factor two. While our tests show that this has little effect on the propagation of the conduction front, it is worth bearing in mind when more complex problems are solved. Also, it should be noted that the mathematically most satisfactory formalism of conduction, involving the variable $u$, cannot be used in implicit numerical schemes. 

\subsection{Heat flux limiting}

Heat flux saturation becomes important in many of our experiments, especially at the most energetic extremes of the heating parameter space that we explore. The flux-saturation (or free-streaming) limit describes the maximum conducted heat flux that the plasma is capable of supporting. A reasonable estimate of this limit can be obtained by assuming that it is reached when all of the particles travel in the same direction at the electron thermal speed. \cite{Bradshaw_2006} give the saturated heat flux as:

\begin{equation}
F_{cs} = \frac{3}{2\sqrt{m_p}} n_p \left(k_B T_p \right)^\frac{3}{2},
\label{eqnA15}
\end{equation}

\noindent where the subscript $p$ denotes the particle species (electrons or ions). More accurate kinetic models (e.g. Fokker-Planck calculations) suggest that Equation~\ref{eqnA15} should be multiplied by a factor $1/6$.  To ensure a smooth transition between heat flux regimes (saturated and unsaturated) in HYDRAD we use the form:

\begin{equation}
F_c = \frac{F_{\mbox{classical}} F_{cs}}{\sqrt{F_{\mbox{classical}}^2 + F_{cs}^2}},
\label{eqnA16}
\end{equation}

\noindent where $F_{\mbox{classical}}$ is the heat flux calculated in the Spitzer-H\"{a}rm approximation (e.g. Equation~\ref{eqnA9} above) and $F_{cs}$ is evaluated at the cell boundary, assuming that the density $n$ remains approximately constant across the grid cell. \cite{West_2008} found that heat flux saturation does not arise in loops that have reached (or are close to) hydrostatic equilibrium, but it does seem essential to include in all dynamic coronal modeling codes.


\end{document}